\documentclass[twocolumn,prd,amssymb]{revtex4}

\newcommand{\be}{\begin{equation}}
\newcommand{\ee}{\end{equation}}
\newcommand{\ba}{\begin{eqnarray}}
\newcommand{\ea}{\end{eqnarray}}
\def\qed{\hbox{${\vcenter{\vbox{
   \hrule height 0.4pt\hbox{\vrule width 0.4pt height 6pt
   \kern5pt\vrule width 0.4pt}\hrule height 0.4pt}}}$}}
\begin{document}
\title{Cosmological Term and Fundamental Physics\footnote{This essay received an
``Honorable Mention'' in the Annual Essay Competition of the Gravity Research 
Foundation for the year 2004.}}

\author{R. Aldrovandi}

\author{J. P. Beltr\'an Almeida}

\author{J. G. Pereira}

\affiliation{Instituto de F\'{\i}sica Te\'orica,
Universidade Estadual Paulista \\
Rua Pamplona 145,
01405-900 S\~ao Paulo, Brazil}

\begin{abstract}
A nonvanishing cosmological term in Einstein's equations implies a nonvanishing spacetime
curvature even in absence of any kind of matter. It would, in consequence, affect  many of
the underlying kinematic tenets of physical theory. The usual commutative spacetime
translations of the Poincar\'e group would be replaced by the mixed conformal translations of
the de Sitter group, leading to obvious alterations in elementary concepts such as time,
energy and momentum. Although negligible at small scales, such modifications may come to have
important consequences both in the large and for the inflationary picture of the early
Universe. A qualitative discussion is presented which suggests deep changes in Hamiltonian,
Quantum and Statistical Mechanics. In the primeval universe as described by the standard
cosmological model, in particular, the equations of state of the matter sources could be
quite different from those usually introduced.
\end{abstract}

\maketitle

\renewcommand{\thesection}{\arabic{section}}
\section{Introduction}

Recent observational data \cite{recent} reveal the presence of a nonvanishing positive
cosmological constant $\Lambda$. Although the astrophysical and cosmological consequences of
this fact have already been extensively studied, an analysis of the consequences for
fundamental Physics is still lacking. The main reason for such absence is, probably,  that
the local effects of a small cosmological term would be negligible and hardly detectable
experimentally. From the conceptual point of view, however, such analysis is not only
justified but highly desirable. In addition, it could find some applications in the cosmology
of the early universe: inflation requires a very high value for $\Lambda$ and the physical
laws applicable in the  primeval universe could eventually be different from those of the
ordinary Physics as we know it. With these motivations in mind, the basic purpose of this
essay will be to make a qualitative discussion on how the presence of a cosmological term
could eventually produce changes in ordinary fundamental Physics.

The crucial consequence of a cosmological term in the sourceless Einstein's equation
\be
R_{\mu \nu} - {\textstyle{\frac{1}{2}}}\, g_{\mu \nu} \, R - \Lambda \, g_{\mu \nu} = 0
\ee
is that Minkowski spacetime, for which $R_{\mu \nu} = 0$, is no more a solution:
contracting with $g^{\mu \nu}$ leads clearly to the scalar curvature  $R = -\, 4\, \Lambda$
and that, inserted back in the equation, to the Ricci tensor $R_{\mu \nu} = -\,
\Lambda \, g_{\mu \nu}$. The solution is a uniformly curved de Sitter
spacetime~\cite{weinberg}. For a positive $\Lambda$ this four-dimensional space can be seen
as a hypersurface in the five-dimensional pseudo-Euclidean space ${\bf E}^{4,1}$ 
with metric $\eta_{AB}$ = $(+1,-1,-1,-1,-1)$, an inclusion whose points in Cartesian
coordinates $(\xi^A) = (\xi^0, \xi^1, \xi^2, \xi^3$, $\xi^{4})$ satisfy~\cite{ellis}
\be
\eta_{AB} \xi^A \xi^B =
-\, L^2.
\label{dshyper}
\ee
The ``pseudo-radius'' $L$ is a length parameter  related to $\Lambda$ by
\be
\Lambda = \frac{3}{L^2}
\label{lambdaR}
\ee
and, consequently, to the curvature.
That hypersurface has the de Sitter group $SO(4,1)$ as group of motions and will be
accordingly denoted $dS(4,1)$. Using $\eta_{a b}$ ($a, b = 0,1,2,3$) for the Lorentz metric
$\eta = $ diag $(1$, $-1$, $-1$, $-1)$, Eq.~(\ref{dshyper}) can be put in the form
\be
\eta_{a b} \, \xi^{a} \xi^{b} - \left(\xi^4\right)^2 = -\, L^2.
\label{dspace2}
\ee
This is the equation defining the de Sitter hypersurface in terms of the five-dimensional
ambient space coordinates $\xi^A$.  It is possible to describe the same hypersurface in
terms of four-dimensional stereographic coordinates.

\section{Stereographic Coordinates}

Such coordinates $x^a$ are obtained through a stereographic
projection from the de Sitter hypersurface (\ref{lambdaR}) into a Minkowski spacetime. When
the Minkowski plane is placed at $\xi^4=0$, the projection is given by~\cite{gursey}
\be
\xi^{a} = h^{a}{}_{\mu} \, x^\mu \equiv \Omega(x) \, \delta^{a}{}_{\mu} \, x^\mu
= \Omega(x) \, x^a
\label{xix}
\ee
and
\be
\xi^4 = - \, L \, \Omega(x) \left(1 +
\frac{\sigma^2}{4 L^2} \right),
\label{xi4}
\ee 
where
\be
\Omega(x) = \frac{1}{1 - \sigma^2 / 4 L^2} \, ,
\label{n}
\ee
with $\sigma^2 = \eta_{a b} \, x^a x^b = \eta_{\mu \nu} \, x^\mu x^\nu$ the
Lorentz invariant interval. In these expressions we have used the relations $x^a =
\delta^{a}{}_{\mu} \, x^\mu$ and $\eta_{\mu \nu} = \delta^{a}{}_{\mu} \,
\delta^{b}{}_{\nu} \, \eta_{a b}$, which means that $\delta^{a}{}_{\mu}$ is a kind of
trivial tetrad. The $h^{a}{}_{\mu}$ introduced in (\ref{xix}), on the other hand, are
components of a nontrivial tetrad field, actually of the 1-form basis members $h^{a} =
h^{a}{}_{\mu} dx^\mu = \Omega(x) \, \delta^{a}{}_{\mu} dx^\mu = \Omega(x) \, dx^a$. In
terms of the stereographic coordinates, the de Sitter line element $ds^2 = \eta_{AB} \,
d\xi^A d\xi^B$ is found to be $ds^2 = g_{\mu \nu} \,d x^\mu dx^\nu$, with
\be
g_{\mu \nu} = h^{a}{}_{\mu} \, h^{b}{}_{\nu} \, \eta_{a b} \equiv \Omega^2(x) \,
\eta_{\mu \nu}.
\label{44}
\ee
The de Sitter space is, therefore, conformally flat, with the conformal factor 
$\Omega^2(x)$. Notice that we are carefully using the Latin alphabet for the algebra
(nonholonomous) indices, and the Greek alphabet for the (holonomous) homogeneous space
fields and cofields. Tangent-space and spacetime indices are, as usual, interchanged with
the help of the tetrad field.

\section{Kinematic Group}

\subsection{Generators}

The kinematic group of any spacetime will always have a subgroup accounting for both the
isotropy of space (rotation group) and the equivalence of inertial frames (boosts). The
remaining transformations, generically called {\it translations}, can be either
commutative or not, and are responsible for homogeneity. The best known relativistic example
is the Poin\-ca\-r\'e group ${\mathcal P}$, naturally associated with the Minkowski spacetime
$M$ as its group of motions. It contains, in the form of a semi-direct product, the Lorentz
group ${\mathcal L} = SO(3,1)$ and the translation group ${\mathcal T}$. The latter acts
transitively on $M$ and its manifold can be identified with $M$. Minkowski spacetime is a
homogeneous space under ${\mathcal P}$, actually the quotient $M \equiv {\mathcal T} = {\cal
P}/{\mathcal L}$. The invariance of $M$ under the transformations of ${\mathcal P}$ reflects
its uniformity. The Lorentz subgroup provides for isotropy around a given point of $M$, and
translation invariance enforces that isotropy around any other point. This is the usual
meaning of ``uniformity", in which ${\mathcal T}$ is responsible for the equivalence of all
points.

Now, $dS(4,1)$ is also a homogeneous spacetime:
\[
dS(4,1) = SO(4,1)/SO(3,1).
\]
The de Sitter group is consequently a bundle with $dS(4,1)$ as base space and the Lorentz
group as fiber \cite{kono}. In the Cartesian coordinates $\xi^A$, the generators of the
infinitesimal de Sitter transformations are
\be
J_{A B} = \eta_{AC} \, \xi^C \, \frac{\partial}{\partial \xi^B} -
\eta_{BC} \, \xi^C \, \frac{\partial}{\partial \xi^A}.
\label{dsgene}
\ee
In terms of the stereographic coordinates $x^a$, these generators are written as
\be
J_{a b} =
\eta_{ac} \, x^c \, P_b - \eta_{bc} \, x^c \, P_a
\label{dslore}
\ee
and
\be
J_{4 a} = L \, P_a - (4 L)^{-1} \, K_a,
\label{dstra}
\ee
where
\be
P_a = \frac{\partial}{\partial x^a} \quad {\rm and} \quad
K_a = \left(2 \eta_{ab} x^b x^c - \sigma^2 \delta_{a}{}^{c}
\right) P_c
\label{cp2} 
\ee
are, respectively, the generators of translations and {\it proper} conformal transformations.
The  $J_{a b}$'s  generate the Lorentz subgroup, whereas the  $J_{4 a}$'s establish the
transitivity of the homogeneous de Sitter space.

\subsection{Transitivity}

The crucial point in the considerations above is the interplay of distinct notions of
space-distance and time-interval: those usual, related to translations, and new ones,
defined by conformal transformations. In fact, as can be seen from the definition of $J_{4 a}$,
the transitivity of the de Sitter spacetime is defined by a mixture of usual translations and
proper conformal transformations. Since $L = (3/\Lambda)^{1/2}$, the relative importance of
each one of these transformations is ultimately determined by the value of the cosmological
constant.

To study the limit of a vanishing cosmological term ($L \to \infty$), it is convenient to
define the generators
\be
\Pi_a \equiv L^{-1} \, J_{4 a} = P_a - (2 L)^{-2} K_a,
\label{l0}
\ee
from where we see that, in this limit, $\Pi_a$ reduces to ordinary translations, and the de
Sitter group reduces to the Poincar\'e group. At the same time, the de Sitter space becomes
the Minkowski spacetime $M$, which is transitive under ordinary translations. On the other
hand, for studying the limit of large values of the cosmological term ($L \to 0$), it is
convenient to define the generators
\be
{\mathcal K}_a \equiv 4 L \, J_{4 a} = 4 L^2 P_a - K_a,
\label{linf}
\ee
from where we see that, in this limit, ${\mathcal K}_a$ reduces to (minus) the proper
conformal generator, and the de Sitter group reduces to the {\em conformal} Poincar\'e group
\cite{ap1}. In this case, the de Sitter space becomes a four-dimensional cone $N$, which is
transitive under proper conformal transformations \cite{aap2}. It is impossible to move
between two arbitrary points of  such a spacetime by ordinary translations, but they can
always be connected by some proper conformal transformation. The cone-space $N$ can be seen
as singular from the point of view of ordinary spacetime translations, but it is smoothly
connected through conformal transformations. 

\section{Exploring the Changes: New Fundamental Physics?}

An immediate consequence of replacing Minkowski by de Sitter as the spacetime underlying
the universe in the presence of a nonvanishing $\Lambda$ concerns the concept of relativistic
field. In Physics as we know it relativistic fields, and the particles which turn up as their
quanta, are classified by the representations of the Poincar\'e group. Each representation is
fixed by the values of two Casimir invariants. As any function of two invariants is also
invariant, it is possible to work with those which  have a clear relationship with simple
physical characteristics: mass and spin. From all the families of representations of the
Poincar\'e group~\cite{Gil74}, Nature seems to have given preference to one of the so-called
discrete series, whose representations are fixed by two invariants  with values
\be 
C_2 = -\, P_a P^a = \qed = -\ m^2 ;\quad  C_4 = -\ m^2 s (s+1).
\ee
The first, involving only translation generators, fixes the mass. It defines the 4-dimensional
Laplacian operator and, in particular, the Klein-Gordon equation
\be 
(\qed + m^2) \phi = 0,
\ee
which all relativistic fields satisfy. The second invariant is the square of the Pauli-Lubanski
operator, used to fix the spin. 

The invariant $C_2$ changes in the presence of a nonvanishing $\Lambda$. The de Sitter group
representations appear in analogous series, one of which tends to the Poincar\'e series above
in the $L \to \infty$ limit, but the corresponding values are \cite{Pol}
\be 
C_2 =   \qed = -\ m^2 + L^{-2} \, [s (s+1) - 2], 
\ee
where now $\qed$ is the Laplace-Beltrami operator on de Sitter space. A scalar ($s= 0$) field
would now obey
\be 
\left[\qed + m^2 - \frac{R}{6} \right] \phi = 0.
\ee
This, by the way, could be the solution to the famous controversy on the $R/6$ factor: the
field $\phi$ above is not a Poincar\'e scalar, but a de Sitter scalar, which means to be
invariant under a transformation including, in addition to Lorentz and translations, also
(proper) conformal transformations. Of course, in the presence of gravitation, $R$ will
represent the total (gravitation plus background) scalar curvature. Mass, in particular, will
be defined according to the field behavior under de Sitter translations, not under Poincar\'e
translations. This change is directly related with the transitivity property of the de Sitter
spacetime, whose generators (appropriate for studying the limit $L \to \infty$) are defined by
$\Pi_a$. It is important to notice that, not only in the case of a scalar field but for any
matter field, the kinetic term of the Lagrangian will necessarily be written with the
``de Sitter derivative'' $\Pi_a$. Notice also that, for studying the limit of large large
values of the cosmological term ($L \to 0$), the corresponding field equations must be
written in terms of the appropriate ``de Sitter derivative'' ${\mathcal K}_a$. 

The same modifications occur in the mechanics of point particles. In fact, the classical
angular momentum of a particle of mass $m$ associated with the de Sitter group is
\be
\lambda_{A B} = m c \left(\xi_A \frac{d \xi_B}{d s} - \xi_B \frac{d \xi_A}{d s} \right),
\ee
with $ds$ the de Sitter line element. In terms of the stereographic coordinates, their
components are written as
\be
\lambda_{\mu \nu} = x_\mu \, p_\nu - x_\nu \, p_\mu
\ee
and
\be
\pi_\mu \equiv L^{-1} \, \lambda_{4 \mu} = p_\mu - (2 L)^{-2} \, k_\mu,
\label{dstra3}
\ee
where $p_\mu=m\,c\,u_\mu$ is the ordinary momentum, and $k_\mu = \bar{\delta}_\mu{}^\nu
\, p_\nu$ is the conformal momentum, with
\be
\bar{\delta}_\mu{}^\nu = \left(2\, \eta_{\mu \rho} \, x^\rho \, x^\lambda -
\sigma^2 \, \delta_\mu{}^\lambda \right) \delta_\lambda{}^\nu
\ee
a kind of {\it conformal} Kr\"onecker delta. The corresponding Hamiltonian is
\be
H \equiv \pi_0 = p_0 - (2 L)^{-2} \; k_0.
\ee
We remark that $\pi_\mu$ is the Noether conserved current under the transformations
generated by $\Pi_a$. In the limit $L \to \infty$, the ordinary notions of momentum and
energy are recovered.

Now, due to the fundamental role played by energy and momentum, we may say that these
modifications will affect all branches of Physics. This is the case, for example, of
Statistical Mechanics. In particular, the equations of state used in the standard cosmological
model for the early Universe should be reexamined. It is not clear, for instance, whether a 
photon gas will retain its usual characteristics in the presence of a cosmological term. 
Analogous changes will also affect Quantum Mechanics. Defining the operator
\be
\hat{p}_\mu = - \, i \hbar \partial_\mu,
\ee
the modified momentum will be
\be
\hat{\pi}_\mu = - \, i \hbar \, \left[\delta_\mu{}^\nu -
(2 L)^{-2} \; \bar{\delta}_\mu{}^\nu \right] \, \partial_\nu.
\ee
It is then easy to verify that
\be
[\hat{\pi}_\mu, \hat{x}^\nu] = -\, i \, \hbar \, \left[\delta_\mu{}^\nu -
(2 L)^{-2}\, \bar{\delta}_\mu{}^\nu\right].
\ee
By considering the nonrelativistic contraction limit $c \to \infty$ \cite{inonu}, which in
this case yields the Newton-Hooke groups \cite{nhg}, the corresponding $\Lambda$-modified
nonrelativistic notions of momentum and energy can be obtained \cite{abcp}. It is then
possible to get the modified version of the Schr\"odinger equation, as well as the
operators commutation rules and the corresponding Heisenberg uncertainty relations. Of course,
the importance of these modifications depends on the value of the cosmological term
$\Lambda$. Accordingly, although they may be negligible locally, they can eventually be
important for the Physics of the very early universe. It should be mentioned that, besides
the problem of adapting the usual formalisms, there are questions of principle, for example,
difficulties concerning the lower bound of the quantum Hamiltonian \cite{Fro74}, possibly
solved by localizability considerations \cite{Mal74}. Anyhow, as said, Quantum Mechanics
itself wll be modified by the presence of $\Lambda$, and should be reconsidered.

\section{Conclusions}

As we have seen, the presence of a cosmological term introduces the conformal generators
in the definition of spacetime transitivity. As a consequence, the conformal transformations
will naturally be incorporated in Physics, and the corresponding conformal current will appear
as part of the Noether conserved current. Of course, for a small enough cosmological term, the
conformal modifications become negligible and ordinary Physics remains valid. For large
values of $\Lambda$, however, the conformal contributions to the physical magnitudes cannot be
disregarded, and these contributions will give rise to deep conceptual changes. This could
be the case, for example, of the early stages of the universe, which according to the
inflationary models, is characterized by a very high value of $\Lambda$. It could also hold
for large enough distances, which are also related to a remote past in the universe history.
In the particular case of extremely large values of $\Lambda$, the ordinary notions of
momentum and energy would become negligible, with only the corresponding conformal notions
surviving. Under such extreme circumstances, a very peculiar new world emerges \cite{aap2},
whose Physics has yet to be developed from the very beginning. Of course, the whole program of
rewriting so much of Physics would be a most daring enterprise. Our point here is only to call
attention to the fact that facing such a program may come to be inevitable. 

\begin{acknowledgments}
The authors would like to thank FAPESP-Brazil, CAPES-Brazil, and CNPq-Brazil for
financial support.
\end{acknowledgments}


\begin{thebibliography}{99}

\bibitem{recent}
See A. G. Riess {\it et al}, Ap. J. {\bf 116},
1009 (1998); S. Perlmutter {\it et al}, Ap. J. {\bf 517}, 565 (1999); P. de Bernardis
{\it et al}, Nature {\bf 404}, 955 (2000) (BOOMERanG); S. Hanany {\it et al}, Ap. J.
Letters {\bf 545}, 5 (2000) (MAXIMA); D. N. Sperger {\it et al}, Ap.J. Supp. {\bf 148}, 175
(2003) [astro-ph/0302209].

\bibitem{weinberg}
S. Weinberg, {\it Gravitation and Cosmology} (Wiley, New York, 1972).

\bibitem{ellis}
S. W. Hawking and G. F. R. Ellis, {\it The Large Scale Structure of Space-Time}
(Cambridge University Press, Cambridge, 1973).

\bibitem{gursey}
F. G\"ursey, {\it Group Theoretical Concepts and Methods in Elementary Particle Physics},
ed. by F. G\"ursey, Istanbul Summer School of Theoretical Physics (Gordon and Breach, New
York, 1962); notice that our notation differs slightly from G\"ursey's notation.

\bibitem{kono}
S. Kobayashi and K. Nomizu, {\it Foundations of Differential Geometry}
(Interscience, New York, 1963).

\bibitem{ap1}
R. Aldrovandi and J. G. Pereira, {\it A Second Poincar\'e Group}, in {\it Topics
in Theoretical Physics: Festschrift for A. H. Zimerman}, ed. by H. Aratyn {\it et al}
(Funda\c c\~ao IFT, S\~ao Paulo, 1998) [gr-qc/9809061].

\bibitem{aap2}
R. Aldrovandi, J. P. Almeida and J. G. Pereira, {\it A Singular Conformal Universe}
[gr-qc/0403099].

\bibitem{Gil74}
E. Wigner, Ann. Math. {\bf 40}, 39 (1939); R.~Gilmore,
{\em Lie Groups, Lie Algebras, and Some of Their Applications}
(Wiley, New York, 1974).

\bibitem{Pol}
See, for example, S. A. Pol'shin, J. Phys. {\bf A33}, 5077 (2000).

\bibitem{inonu}
E. In\"on\"u and E. P. Wigner, Proc. Natl. Acad. Scien. {\bf 39}, 510 (1953).

\bibitem{nhg}
H. Bacry and J.-M. L\'evy-Leblond, J. Math. Phys. {\bf 9}, 1605 (1968);
C. Duval, G. Burdet, H. P. K\"unsle and M. Perrin, Phys. Rev. D {\bf 31}, 1841 (1985).

\bibitem{abcp}
R. Aldrovandi, A. L. Barbosa, L. C. B. Crispino and J. G. Pereira, Class. Quant. Grav. {\bf
16}, 495 (1999) [gr-qc/9801100];
G. W. Gibbons and C. E. Patricot, Class. Quant. Grav. {\bf 20}, 5225 (2003) [hep-th/0308200].

\bibitem{Fro74}
C. Fronsdal, Phys. Rev. {\bf D10}, 589 (1974), and references therein.

\bibitem{Mal74}
S. Malin, Phys. Rev. {\bf D9}, 3228 (1974); T. O. Philips and E. P. Wigner, in {\em Group
Theory and its Applications}, ed. by E. M. Loebl (Academic Press, New York, 1968).

\end{thebibliography}
\end{document}